\documentclass[a4paper,fleqn]{cas-dc}

\usepackage[numbers]{natbib}
\usepackage{bm}

\def\tsc#1{\csdef{#1}{\textsc{\lowercase{#1}}\xspace}}
\tsc{WGM}
\tsc{QE}
\tsc{EP}
\tsc{PMS}
\tsc{BEC}
\tsc{DE}

\begin{document}
\shorttitle{}
\shortauthors{T. Kanao et~al.}

\title [mode = title]{Layer-selective detection of magnetization directions from two layers of antiferromagnetically-coupled magnetizations by ferromagnetic resonance using a spin-torque oscillator}                      



\author{Taro Kanao}
\ead{taro.kanao@toshiba.co.jp}
\address{Corporate Research and Development Center, Toshiba Corporation, 1, Komukai-Toshiba-cho, Saiwai-ku, Kawasaki 212-8582, Japan}
\author{Hirofumi Suto}
\author{Koichi Mizushima}
\author{Rie Sato}

\cortext[cor1]{Corresponding author}


\begin{abstract}
We use micromagnetic simulation to demonstrate layer-selective detection of magnetization directions from magnetic dots having two recording layers by using a spin-torque oscillator (STO) as a read device. This method is based on ferromagnetic resonance (FMR) excitation of recording-layer magnetizations by the microwave field from the STO. The FMR excitation affects the oscillation of the STO, which is utilized to sense the magnetization states in a recording layer. The recording layers are designed to have different FMR frequencies so that the FMR excitation is selectively induced by tuning the oscillation frequency of the STO. Since all magnetic layers interact with each other through dipolar fields, unnecessary interlayer interferences can occur, which are suppressed by designing magnetic properties of the layers. We move the STO over the magnetic dots, which models a read head moving over recording media, and show that changes in the STO oscillation occur on the one-nanosecond timescale.
\end{abstract}



\begin{keywords}
Three-dimensional magnetic recording \sep
Spin-torque oscillator \sep
Ferromagnetic resonance \sep
Micromagnetic simulation
\end{keywords}

\maketitle
\section{Introduction}\label{sec_intro}
Ferromagnetic resonance (FMR) has long been studied as a subject of fundamental physics as well as a tool for observing the magnetic states of materials and devices.
Recently, the use of FMR has been proposed for three-dimensional (3D) (or multilayer) recording to increase recording density in hard disk drives (HDDs)~[1--10].
The use of FMR enables layer-selective writing~[1--7]
and reading~\cite{Yang2013, Yang2013a} by employing recording layers (RLs) with different FMR frequencies and then selectively inducing FMR by using a microwave field of the corresponding frequency.

In magnetic recording, it is necessary to manipulate and sense nanometer-sized magnetizations within the nanosecond timescale.
For this kind of nanometer- and nanosecond-scale FMR excitation and measurement, a spin-torque oscillator (STO)~[11--14]
can be used~\cite{Suto2014, Suto2017}, which is a nanometer-sized microwave field generator with a typical oscillation frequency of approximately $0.1$ GHz to several tens of gigahertz.
In write operations, an RL is selectively switched by the combination of the microwave field from the STO and a field from a write pole, which is a multilayer version of microwave-assisted magnetization switching~[17--20].

Read operations, which are the focus of this paper, utilize changes in the STO oscillation when the STO induces FMR in an RL.
This read method is therefore called resonance reading.
The basic principles have been experimentally demonstrated~\cite{Suto2014}.
An STO suitable for resonance reading in HDDs has been proposed, and its transient magnetization dynamics during the resonance reading for the case of a single recording layer have been investigated by using micromagnetic simulation~\cite{Kanao2018}.
Envelope detection has been applied as the signal processing method for the waveform of the STO~\cite{Nakamura2018, Nakamura2018a}.

Resonance reading detects the FMR of a single layer selectively, and does not sense superposition of stray fields from each layer.
Thus, resonance reading can be applied to antiferromagnetically-coupled (AFC) media where the stray fields are suppressed~[24--27].
The use of this kind of AFC media has been proposed for 3D recording to avoid shifts in the FMR frequencies by interlayer stray fields~\cite{Yang2013a, Suto2016a}.
However, designs for fully layer-selective resonance reading, namely, large read signals from each layer and no effect from the other layer, have not been reported~\cite{Kanao2018a}.

\begin{figure}
	\includegraphics[width=8.5cm,bb=0 0 503 446]{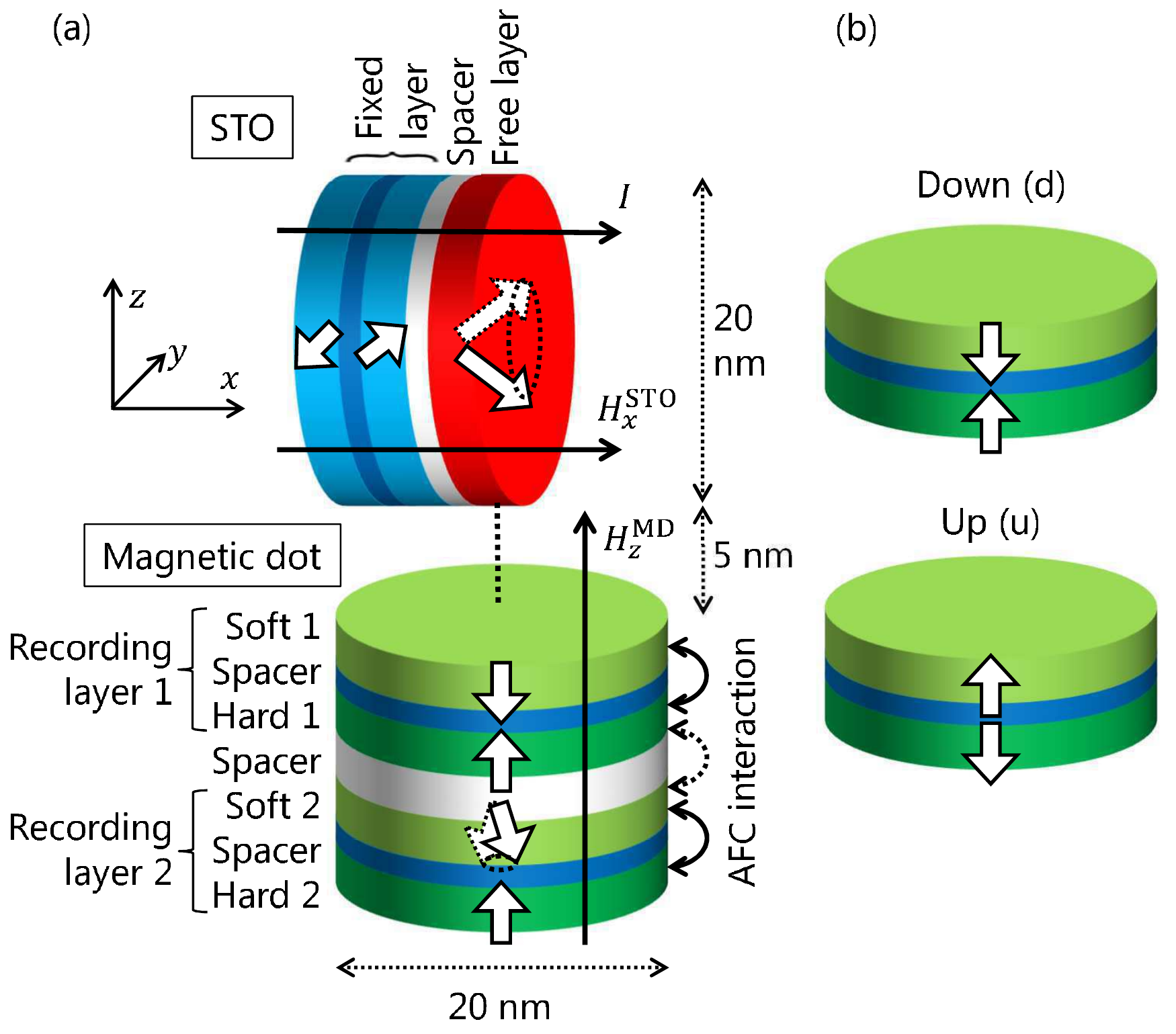}
	\caption{(a) Schematic of a spin-torque oscillator (STO) and a magnetic dot (MD) with two antiferromagnetically-coupled (AFC) recording layers (RLs).
		(b) Down and up states of an AFC-RL.}
	\label{fig_sto_dot}
\end{figure}

In this paper, we use micromagnetic simulation to demonstrate layer-selective resonance reading from vertically stacked double-layer recording media.
Figure~\ref{fig_sto_dot}(a) shows a schematic diagram of an STO and a magnetic dot (MD) with double RLs.
We find that a straightforward stacking of two AFC layers fails to allow the layer-selective reading because an AFC structure cannot cancel the stray field around it.
To solve this problem, we carefully design the magnetic properties of each layer in the MD.
As a result, we show that inducing FMR excitation in a layer reduces the STO oscillation power, and different layers can be selected by changing the STO oscillation frequency.
Furthermore, we move the STO over MDs, and demonstrate changes in the STO oscillation on the one-nanosecond timescale.

\section{Model design}\label{sec_model}
\subsection{Structure}
We now look at Fig. \ref{fig_sto_dot}(a) in more detail.
We use an STO with a perpendicular free layer and in-plane fixed layers~\cite{Kanao2018}; this STO exhibits stable magnetization oscillation with large output power~\cite{Kubota2013, Taniguchi2013} high enough for read sensor applications in HDDs~[31--33].
A current $I$ and an external field $H^{\rm STO}_x$ are applied in the stacking direction and these are used to tune the STO oscillation power and frequency, which enables the STO to access each RL.
Note that read operations performed by preparing two STOs to access two RLs are possible and are expected to improve read performance~\cite{Nakamura2018a}.
In that case, tuning of the STO oscillation frequency by the external field is not necessary, and bias-field free STOs~\cite{Zeng2013a} can be utilized.
In this work, however, we assume a single STO for simplicity.

The MD has two AFC-RLs, each of which consists of a soft layer (SL) and a hard layer (HL)~\cite{Yang2013a, Suto2016a}.
We refer to the upper layer as RL1 and the lower layer as RL2.
The SLs and HLs have small and large perpendicular anisotropies, respectively.
The SLs have FMR frequencies near the STO oscillation frequency and are used for reading.
The HLs have high thermal stability and store the written data.
The FMR frequencies of the HLs are much higher than the SLs, and the HLs are excited only a little during read operations~\cite{Kanao2018}.
Thus, possibility of switching the HLs during read operations (unintentional data erasure) can be prevented.

Each RL has two AFC states as shown in Fig.~\ref{fig_sto_dot}(b).
The AFC state is spontaneously obtained when effect of the antiferromagnetic interlayer coupling $J_{\rm ex}$ is larger than that of the perpendicular anisotropy of the SL and the stray field from the HL.
Under this condition, the magnetization of the SL is in the direction opposite to that of HL in stable states.
We refer to these two AFC states as down and up according to the magnetization direction of the SL.
These AFC states suppress total static stray field.
In all the simulations, an external field ${H^{\rm MD}_z=0.2}$ kOe is applied perpendicularly to the MD such that the FMR frequency depends on the magnetization direction.
We apply $H^{\rm STO}_x$ and $H^{\rm MD}_z$ independently for simplicity of the analysis.
In realistic designs, these fields can be generated by a single source such as a magnetic pole.
The STO and MD are columns with a diameter of 20 nm, and are separated by 5 nm.
The position of the STO is fixed with the free layer above the center of the MD, or moving past the MDs.

\subsection{Parameters}
For the STO, typical parameters of STOs based on tunnel-magnetoresistive films are used~\cite{Suto2017, Kubota2013, Nagasawa2014}.
The values are as follows: free layer thickness 3 nm, Gilbert damping ${\alpha=0.01}$, saturation magnetization ${M_s=1.2}$ kemu/cm$^3$, perpendicular anisotropy ${K_u=5.5}$ Merg/cm$^3$, fixed layer thickness 2 nm, ${M_s=0.8}$ kemu/cm$^3$, spacer thicknesses 1 nm, and spin polarization ${P=0.65}$.
The exchange stiffness is ${A=1.6}$ $\mu$erg/cm for all magnetizations.
To reduce the threshold current of the STO, the magnetization of the fixed layer next to the free layer is tilted 5$^\circ$ with respect to the $y$-direction~\cite{Taniguchi2013, Houssameddine2007}, ${(\sin5^\circ, \cos5^\circ, 0)}$.
This tilt is induced by the $H^{\rm{STO}}_x$.
The magnetization of the other fixed layer is fixed in $-y$-direction for simplicity.

\begin{table}
	\caption{Parameters of magnetic dot.
		Units are thickness (nm), $M_s$ (kemu/cm$^{3}$), $K_u$ (Merg/cm$^{3}$), and $J_{\rm ex}$ (erg/cm$^2$).}
	\label{tab_parameters_dot}
	\begin{center}
		\begin{tabular}{ccccccc}
			\hline\hline
			Layer&&Thickness&$\alpha$&$M_s$&$K_u$&$J_{\rm ex}$\\
			\hline
			RL1&SL1&2&0.01&0.8&3.5&---\\
			&Spacer&1&---&---&---&$-1.0$\\
			&HL1&2&0.02&0.8&9.0&---\\
			\hline
			Spacer&&2&---&---&---&$-0.074$\\
			\hline
			RL2&SL2&2&0.01&1.2&8.1&---\\
			&Spacer&1&---&---&---&$-1.0$\\
			&HL2&2&0.02&2.1&29.0&---\\
			\hline\hline
		\end{tabular}
	\end{center}
\end{table}

Table \ref{tab_parameters_dot} shows the parameters for the MD, which are chosen from ranges of typical values in magnetic materials.
At this point, the following two considerations are required.
The first is that the changes in the STO for the FMR excitation in SL2 should be large, since the interaction is smaller for SL2 owing to the larger distance from the STO.
Thus $M_s$ is set to be larger for SL2 than SL1.
In addition, $K_u$ is chosen such that the effective perpendicular field is smaller for SL2 than for SL1.
The smaller effective perpendicular field leads to larger excitation by the microwave field from the STO~\cite{Okamoto2015, Suto2015a}.

The second consideration is that effects from the RL that is not being read should be avoided.
Note that the stray field near an AFC-RL is not zero even if the SL and HL have the same magnetic volume.
Since the stray field is perpendicular to the other RL, its FMR frequency is sensitive to the stray field.
Thus, the effects of the stray fields among the RLs should be carefully canceled.
We cancel the stray field at SL1 from RL2, by using larger $M_s$ for HL2 than SL2, because HL2 is farther from SL1 than SL2.
On the other hand, we choose the same $M_s$ for SL1 and HL1, and cancel the effect of the stray field at SL2 from RL1 by introducing an antiferromagnetic coupling between HL1 and SL2.
The purpose of the same $M_s$ for SL1 and HL1 is to reduce effects of the stray field on the STO.
If larger $M_s$ for SL1 than HL1 is chosen to cancel the stray field at SL2, then the stray field at the STO becomes strong, which means that the state of RL1 always affects the STO oscillation, preventing the layer-selective reading.
Although the same values of $M_s$ for SL1 and HL1 cause nonzero stray field at the STO, complete cancellation is not required because the STO oscillation is rather robust against the stray field orthogonal to the oscillation axis.
Here, though $M_s$ of HL2 is larger than realistic material values, we use this value as a model and we expect that similar situation can be realized by thicker HL2 or smaller $M_s$ of soft layers.

The value of $\alpha$ is set larger for the HLs than for the SLs in order to suppress magnetization excitations in the HLs.
The $K_u$ of the HLs is high enough for thermal stability, which is proportional to the effective perpendicular anisotropy $K^{\rm eff}_u$~\cite{Okamoto2012b}.
The larger $K_u$ is necessary for HL2, because $K^{\rm eff}_u$ decreases with $M_s$ owing to the demagnetization field.
(Thermal fluctuations of the magnetizations are not treated in this work.)

Please note that in write operations, the effective perpendicular anisotropy field ${H^{\rm eff}_K=2K^{\rm eff}_u/M_s}$ determines the magnetization dynamics during switching~\cite{Okamoto2015, Suto2015a}.
Hence, for designs that include write operations, the values of $K_u$ should be chosen so that $K^{\rm eff}_u$ and $H^{\rm eff}_K$ meet the conditions for thermal stability and switching, respectively.

\subsection{Simulation methods}
The micromagnetic simulations are based on the Landau-Lifshitz-Gilbert-Slonczewski (LLGS) equations for normalized magnetization vectors $\bm{m}_l$ of the magnetic layers $l$
\begin{eqnarray}
	\frac{\partial \bm{m}_l}{\partial t}=-\gamma\bm{m}_l\times\bm{H}^{\rm eff}_l+\alpha_l\bm{m}_l\times\frac{\partial \bm{m}_l}{\partial t}+\bm{T}^S_l,\nonumber
\end{eqnarray}
where $t$, $\gamma$, $\bm{H}^{\rm eff}_l$, and $\bm{T}^S_l$ are the time, gyromagnetic ratio, effective field, and spin-torque term, respectively~\cite{Bertotti2009}.
The $\bm{H}^{\rm eff}_l$ consists of the bias field $\bm{H}^{\rm bias}_l$, exchange field $\bm{H}^{\rm exch}_l$, anisotropy field $\bm{H}^{\rm ani}_l$, dipolar field $\bm{H}^{\rm dip}_l$, and interlayer field $\bm{H}^{\rm int}_l$.
Owing to the $\bm{H}^{\rm dip}_l$ term, all of the magnetizations interact.
The $\bm{H}^{\rm int}_l$ term originates from the Ruderman-Kittel-Kasuya-Yosida (RKKY) interlayer exchange coupling $J_{\rm ex}$~\cite{Parkin1990}.
The LLGS equations are solved by using the finite differential method with 1-nm cubic cells~\cite{Kudo2015, Kudo2010a}.

\section{Simulation results}\label{sec_results}
Under the following conditions, the magnetization of the STO free layer exhibits out-of-plane precession with the oscillation frequency around $10$ GHz, and the magnetizations of the SLs oscillate such that the AFC configurations are maintained.
Since the magnetizations remain almost uniform in these dynamics, we represent the magnetization states by the spatial average of the normalized magnetization vectors.
In particular, we focus on the $y$-component because this component of the STO free layer serves as an output signal for the magnetoresistive effect owing to the magnetization of the fixed layer in the $y$-direction.

\subsection{Layer-selective FMR excitation by STO}
\begin{figure}
	\includegraphics[width=8.5cm,bb=0 0 244 319]{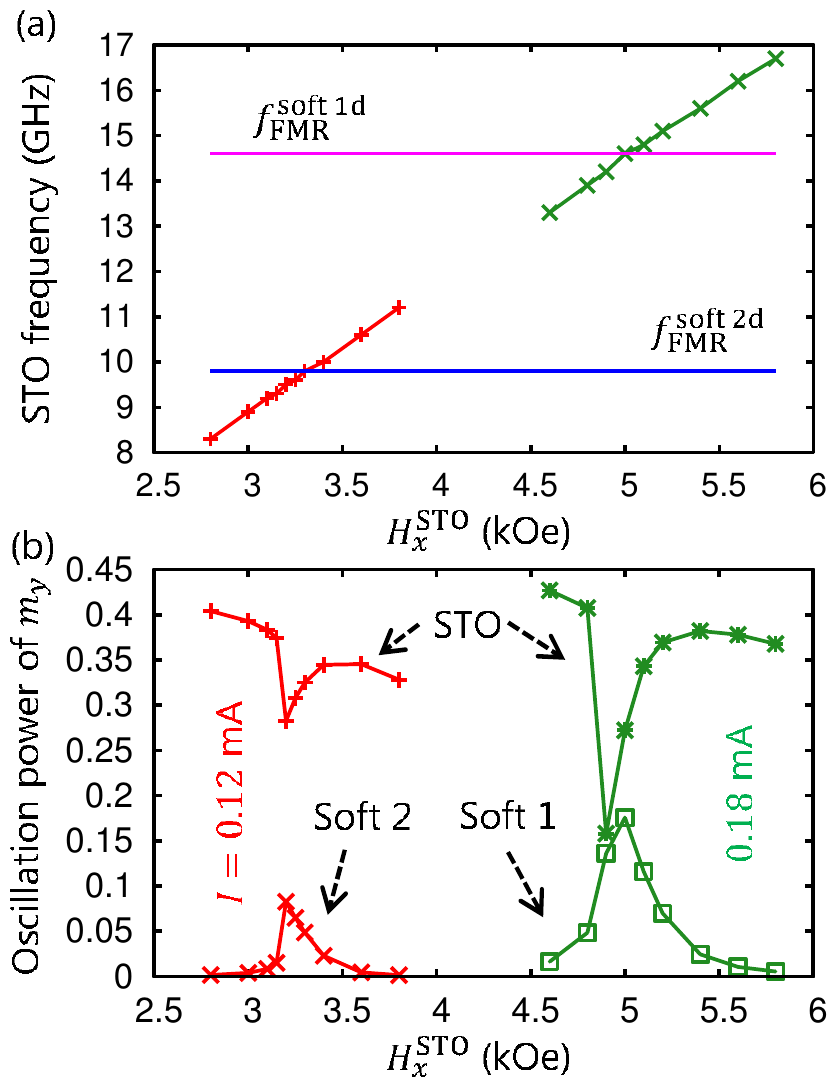}
	\caption{(a) STO oscillation frequency and (b) power of $m_y$ of the STO free layer and the SLs of the MD as a function of $H^{\rm STO}_x$ for the down states of the RLs.
		The cases of ${I=0.18}$ mA and $0.12$ mA are shown for SL1 and SL2, respectively.}
	\label{fig_freq_pow_ab}
\end{figure}

First, we show layer-selective FMR excitation by the STO and the corresponding changes in the STO oscillation.
Figure \ref{fig_freq_pow_ab} shows the STO oscillation frequency and the oscillation power of $m_y$ of the STO free layer and the SLs of the MD as a function of $H^{\rm STO}_x$.
Time evolutions of length 20 ns are calculated and the last 10 ns are used for analysis.
The oscillation power is obtained by ${\overline{\left[m_y-\overline{m_y}\right]^2}}$, where the bars mean time average.

As shown in Fig.~\ref{fig_freq_pow_ab}(a), the STO oscillation frequency increases with increasing $H^{\rm STO}_x$.
The linear dependence is expected from that of the FMR frequency of the perpendicular magnetization, as the corresponding precession mode is excited in the STO by the spin-torque~\cite{Kiselev2004}.
Here, the data are split into two parts because these parts are obtained under different currents $I$, which is explained later.
When the STO oscillation frequency is near the FMR frequency of each SL in ${H^{\rm MD}_z=0.2}$ kOe, $f^{\rm soft 1d}_{\rm FMR}$ and $f^{\rm soft 2d}_{\rm FMR}$, FMR excitation is induced in the SL and the STO oscillation power decreases as in Fig.~\ref{fig_freq_pow_ab}(b).
This decrease is due to dissipation by the FMR excitation in the SL.
The FMR ranges are separated for SL1 and SL2, and each layer can be selectively accessed.

The currents used to access SL1 and SL2 are $I=0.18$ mA and $0.12$ mA, respectively.
The $I$ for SL2 is optimized so that the relative change in STO oscillation power becomes largest.
The reason for the optimal value of $I$ is the following.
When $I$ is too small, the oscillation amplitude of the STO is small and cannot excite the large oscillation in the SL enough for the change in the STO oscillation.
On the other hand, when $I$ is too large, trajectory of the STO oscillation becomes so rigid~\cite{Slavin2009} that the dissipation by the FMR excitation in the SL cannot affect the STO oscillation.
Hence, the relative change in the STO oscillation is maximized at the optimal $I$.
The width of $H^{\rm STO}_x$ where the STO oscillation power significantly responds is about $0.2$ kOe.
This range can be separated for the up and down states of the RLs by ${H^{\rm MD}_z=0.2}$ kOe as shown in the following simulations.

\subsection{Instantaneous power of the STO during resonance reading}
\begin{figure}
	\includegraphics[width=8.5cm,bb=0 0 234 413]{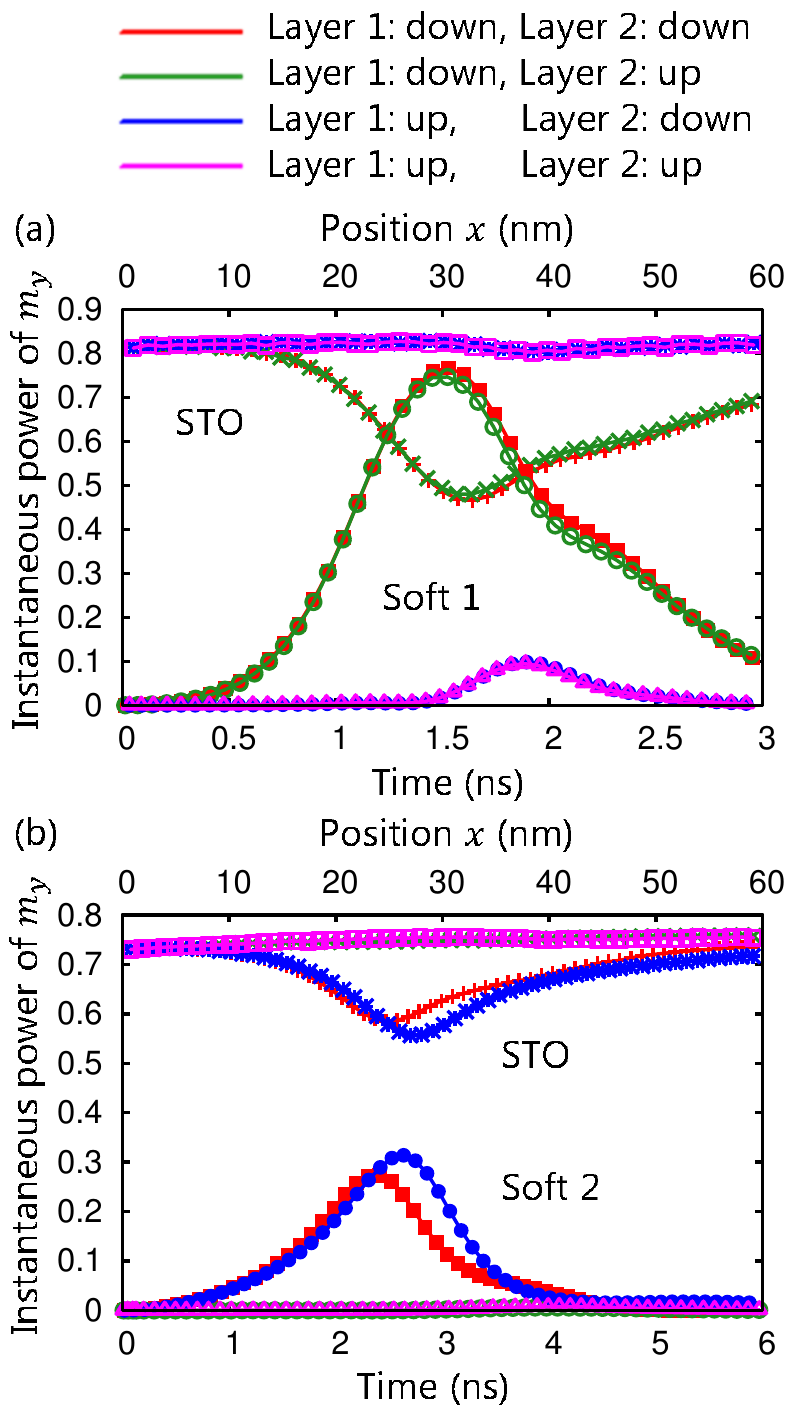}
	\caption{Instantaneous power of $m_y$ of the STO free layer and the SLs of the MD for the 4 states of the MD as a function of time.
		The STO is moved over the MD.
		The position $x$ is indicated.
		(a) ${I=0.18}$ mA, ${H^{\rm STO}_x=5.0}$ kOe, and ${V=20}$ m/s.
		(b) ${I=0.12}$ mA, ${H^{\rm STO}_x=3.25}$ kOe, and ${V=10}$ m/s. 		
		}
	\label{fig_inspow_abdu}
\end{figure}

We now show the magnetization-direction-dependent FMR excitation of each RL and the corresponding time evolution of the STO.
In this case, the STO is moved past the MD.
The STO starts at $20$ nm away from the MD, moves over the center of the MD at velocity $V$, and then stops at $20$ nm away.
Figure~\ref{fig_inspow_abdu} shows the time evolutions of instantaneous power of $m_y$ of the STO free layer and the SLs.
The position $x$ of the STO is also indicated.
The instantaneous power is given by the square of the instantaneous amplitude which is estimated as half of the difference between the maximum and minimum values of $m_y$ in each oscillation period.

In Fig.~\ref{fig_inspow_abdu}(a) and (b), $I$ and $H^{\rm STO}_x$ are set so that FMR excitation is induced in SL1 and SL2, respectively, under ${H^{\rm MD}_z=0.2}$ kOe with both in the down state.
For SL1 and SL2, ${V=20}$ m/s and $10$ m/s are used, respectively, for the following reason.
Since SL2 is farther from the STO, the coupling by the dipolar fields is smaller.
Then, the time necessary for transition to a coupled oscillation state becomes longer~\cite{Slavin2009, Kanao2018}, which means that longer time is necessary for the reading.
We thus use the lower $V$ for SL2.

Figure~\ref{fig_inspow_abdu}(a) shows that for RL1 in the down state, SL1 is resonantly excited and the STO instantaneous power decreases as the STO approaches the MD.
By contrast, for RL1 in the up state SL1 is little excited, which is because the $H^{\rm MD}_z$ shifts the FMR frequency of the SL1 in the up state away from the STO oscillation frequency and FMR excitation does not occur.
The STO instantaneous power then remains almost constant.
Identical results are obtained for the opposite state of RL2.
A shoulder is observed in between $2$ ns and $2.5$ ns, which is thought to reflect distribution of the microwave field around the STO~\cite{Zhu2010}.

In Fig.~\ref{fig_inspow_abdu}(b), despite the longer distance and the presence of RL1 in between, we obtain the similar layer-selective and magnetization-direction-dependent FMR excitation in SL2 and the significant response of the STO, which can be detected depending on the signal-to-noise ratio~\cite{Nakamura2018, Nakamura2018a}.
As the STO is above the MD only during the time of $1$ ns or $2$ ns in Fig.~\ref{fig_inspow_abdu}, these results demonstrate the STO response on one-nanosecond timescale.

In this study, the MD in down state is detected.
For the up state, almost similar response of the STO can be obtained when the frequencies are matched by changing $H^{\rm STO}_x$ or $H^{\rm MD}_z$.
A possible difference is the position (or timing) of the largest resonance, that is, the peaks appear earlier for the down state than for up, which can be seen in Fig.~\ref{fig_inspow_abdu}(a). 
This difference is due to spatial distribution of the rotation direction of the dipolar field from the STO free layer~\cite{Kanao2018, Zhu2010}.

\subsection{Waveforms of resonance reading from 6 MDs}
\begin{figure}
	\includegraphics[width=8.5cm,bb=0 0 251 510]{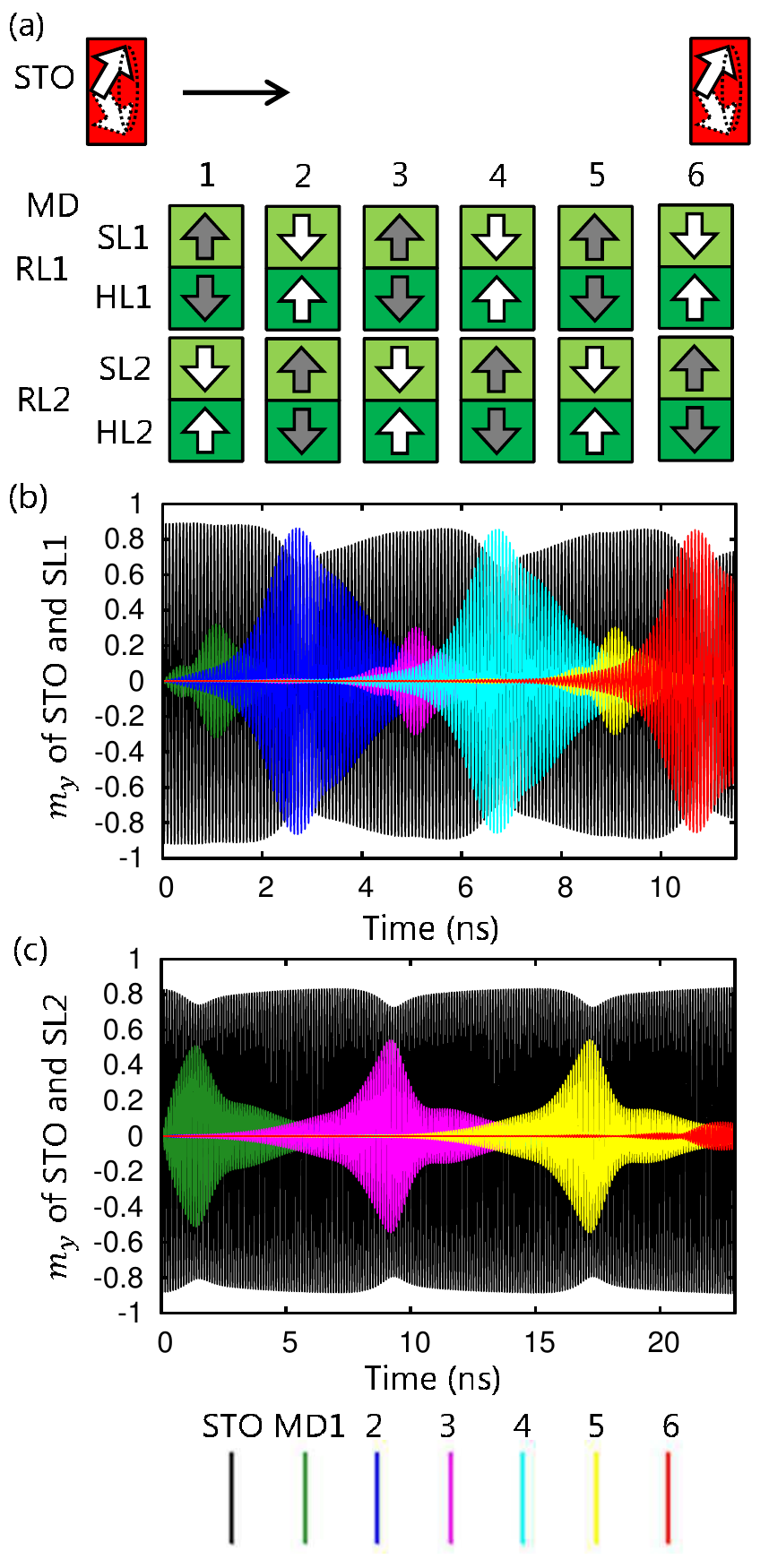}
	\caption{(a) Schematic of the STO moving over 6 MDs.
	Magnetization directions of the MDs are shown.
	(b) Waveforms of $m_y$ of the STO free layer and the SL1s for $I=0.18$ mA, $H^{\rm STO}_x=5.0$ kOe, and $V=20$ m/s, and (c) those of the STO free layer and the SL2s for $I=0.12$ mA, $H^{\rm STO}_x=3.25$ kOe, and $V=10$ m/s.}
	\label{fig_waveform}
\end{figure}

Finally, we show the waveforms of resonance reading from 6 MDs in a row separated by $20$ nm.
The magnetization configurations of the RLs are alternating with adjacent RL1 and RL2 in opposite directions, as shown in Fig.~\ref{fig_waveform}(a).
Fig.~\ref{fig_waveform}(b) and (c) show the waveforms of $m_y$ of the STO free layer and the SL1s and SL2s.
The values of $I$ and $H^{\rm STO}_x$ are set so that FMR excitation is induced in the SL1s in Fig.~\ref{fig_waveform}(b) and the SL2s in Fig.~\ref{fig_waveform}(c) under ${H^{\rm MD}_z=0.2}$ kOe both in the down state.
The STO oscillation amplitude decreases when the STO is near these SLs, and recovers when the STO leaves.
These waveforms of the STO demonstrate the feasibility of nanosecond- and nanometer-scale layer-selective resonance reading.

\section{Discussion}
We have chosen the STO with the modest perpendicular anisotropy in order to reduce complexity of the magnetization dynamics, as follows.
As both the STO and the SL exhibit the out-of-plane precession, the resonance leads to a decrease in the oscillation power (or precession angle) of the STO and an increase in the SL.
The changes in the precession angles cause those of the anisotropy fields, shifting the frequencies.
Thus, when both the STO and the SL have the perpendicular anisotropy, the resonance induces frequency shifts in opposite directions, which stabilizes the coupled oscillations~\cite{Kanao2018}.
On the other hand, an STO with an in-plane anisotropy shows an enhanced interaction with perpendicular magnetic dots, even leading to switching~\cite{Kudo2015, Suto2018}.
We expect that this kind of nonlinearity~\cite{Slavin2009, Kudo2009} can be utilized to improve the resonance reading.

Thermal fluctuations are not included in the simulations.
The thermal fluctuations in the STO cause oscillation linewidth and can degrade signal-to-noise ratio of the read signals~\cite{Mizushima2010, Quinsat2010, Kanao2016}.
This issue is an important future work.
On the other hand, we think that effects of the thermal fluctuations on stability of the MDs are limited, because high enough $K_u$ are used.

In this study, the parameters were optimized for a single MD.
Optimization for the MDs with smaller separation is left as future work in which the stray fields between the MDs are expected to affect the FMR excitation.

\section{Conclusions}\label{sec_summary}
We have demonstrated the layer-selective resonance reading using the STO from the MDs containing the double AFC-RLs by the micromagnetic simulation.
The STO has the perpendicular free layer and the in-plane fixed layers, and the oscillation frequency tuned to around $10$ GHz.
The layer selectiveness can be obtained by the parameters for the RLs such that the change in the STO for RL2 is large and the reading waveform from an RL is unaffected by the magnetization states of the other RL.
The STO oscillation amplitude changes on the one-nanosecond timescale when the STO induces the FMR excitation in an RL, which can be used for detecting the magnetization state.

\section*{Acknowledgments}
We thank S. Okamoto, M. Igarashi, Y. Nakamura, Y. Okamoto, T. Nagasawa, and K. Kudo for valuable discussions.
This work was supported by the Strategic Promotion of Innovative Research and Development from the Japan Science and Technology Agency (JST).


\bibliographystyle{model1-num-names}

\bibliography{mag}

\end{document}